\documentclass[reprint]{revtex4-2}

\usepackage{graphicx}
\usepackage{subcaption}
\usepackage{float}
\usepackage{mathtools}
\usepackage{tabularx}
\usepackage[british]{babel}
\usepackage{csquotes}
\usepackage{dsfont}
\usepackage{textcomp, gensymb}
\usepackage{gensymb}
\usepackage{amsmath}
\usepackage{orcidlink} 
\usepackage{appendix}

\begin{document}

\title{Data-driven Approach for Interpolation of Sparse Data}

\author{{R.F.~Ferguson}
\orcidlink{0009-0004-5853-9671}}
\email[Corresponding author: ]{r.ferguson.3@research.gla.ac.uk}

\author{{D.G.~Ireland}
\orcidlink{0000-0001-7713-7011}}

\author{{B.~McKinnon}
\orcidlink{0000-0002-5550-0980}}

\affiliation{University of Glasgow, Glasgow, G12 8QQ, United Kingdom}

\date{May 2025}

\begin{abstract}
\begin{description}
    \item[Background]  Studies of hadron resonances and their properties are limited by the accuracy and consistency of measured datasets, which can originate from many different experiments. 
    \item[Purpose] We have used Gaussian Processes (GP) to build interpolated datasets, including quantification of uncertainties, so that data from different sources can be used in model fitting without the need for arbitrary weighting. 
    \item[Method]  GPs predict values and uncertainties of observables at any kinematic point. Bayesian inference is used to optimise the hyperparameters of the GP model. 
    \item[Results] We demonstrate that the GP successfully interpolates data with quantified uncertainties by comparison with generated pseudodata. We also show that this methodology can be used to investigate the consistency of data from different sources. 
    \item[Conclusions]  GPs provide a robust, model-independent method for interpolating typical datasets used in hadron resonance studies, removing the limitations of arbitrary weighting in sparse datasets. 
\end{description}
\end{abstract}

\maketitle

\section{Introduction}\label{sec:intro}

Extracting information about hadron resonances requires fitting theoretical models to experimental data. However, this data often comes from different experiments of different physics quantities in varying kinematic regions; studying coupled channels with different kinematic coverages and binning can make direct comparison challenging. The consistency of these datasets directly impacts the quality of the fit, thus making it difficult to accurately constrain the theoretical models. Sparse datasets in key kinematic regions further complicates the quantification of uncertainties, often requiring arbitrary weighting that may introduce bias. 

A robust approach to solving these problems involves utilising Gaussian Processes (GPs), a Bayesian inference machine learning technique that provides probabilistic predictions for unknown datapoints. Unlike traditional machine learning methods, GPs do not require any training; instead, they operate on three fundamental assumptions:
\begin{enumerate}
    \item Some kernel function can be defined to measure the covariance between known datapoints; 
    \item This same kernel function can be used to predict the covariance between unknown datapoints; 
    \item Some idea of the form of the posterior distribution is known (e.g. smoothness, continuity, periodicity). 
\end{enumerate}
This framework enables GPs to interpolate datasets and provide predicted values and uncertainties, which ensures that the quantification of uncertainties is well-defined. Furthermore, the full dataset can be utilised without the need for arbitrary splitting into training and testing sets, avoiding unnecessary loss of data.

To optimise the GP presented here, a Bayesian inference technique is used for hyperparameter tuning, ensuring that the prediction best represents the underlying structure of the data. Validation and testing are performed using pseudodata generated from combinations of Legendre polynomials. This ensures the GP produces unbiased and accurate predictions without the limitations of arbitrary weighting. 

Beyond interpolation, GPs provide a powerful tool for statistical validation of datasets. By comparing measurements from different experiments, inconsistencies, such as variations in coverage or binning, can be identified and addressed. This is particularly useful in coupled-channels analyses, where different datasets must be combined despite their differences in kinematic coverage and binning. Additionally, theorists can use GP built datasets to test models and highlight kinematic regions where theoretical models diverge from the GP predicted empirical trends. By providing a model-independent interpolation method, GPs offer a significant advantage in hadron spectroscopy, improving dataset consistency and enhancing the precision of theoretical analyses.

\section{An Overview of the Gaussian Process}\label{sec:GP_overview}
This is a brief overview of the mathematical principles behind a Gaussian Process fit; for a more in-depth look, Rasmussen and Williams \cite{rasmussen_gp} is widely regarded as the ``standard text" and offers a comprehensive study of the topic. 

Assume that there are \textit{n} known datapoints $(\Vec{x}_i,y_i \pm e_i),\,i=1...n$. Here, $\Vec{x}_i$ is a vector of length $p$ whose parameters are kinematic variables such as energy or scattering angle; it used to define the expression $\Vec{y}=f(X)$ where $\Vec{y}$ is a column vector of length $n$ made up of the scalars $y_i$ (which are the physics quantities of interest) and $X$ is an $n \times p$ matrix whose rows are the collection of $\Vec{x}_i$. Note that $X$ will always be a 2D matrix, regardless of the number of kinematic dimensions. 

Assume that $\Vec{y}$ is drawn from a multivariate Gaussian of the form 
\begin{equation*}
    P(\Vec{y} \mid X) \sim \mathcal{N}(\Vec{0},K),
\end{equation*}
where $K=\kappa(X,X) + \Vec{e}^2\mathbf{I}_n$ is the covariance matrix and $\kappa$ is some kernel function used to measure the covariance between two vectors. Note that by assuming the multivariate Gaussian has a zero mean does not impact the prediction and only simplifies some derivation later. Here $K_{ab}=\kappa(\Vec{x}_a,\Vec{x}_b)+\delta_{ab}e_a^2 \text{ where } \Vec{x}_a,\Vec{x}_b$ are rows of the matrix $X$. 

Assume there are \textit{m} datapoints of the form outlined previously, with known $\Vec{x}_{*j}$ and unknown $y_{*j},e_{*j}$, which are correlated to the \textit{n} known datapoints. Note that $\forall i, \exists j$ such that $\Vec{x}_i=\Vec{x}_{*j}$. An $m \times p$ matrix $X_*$ can then be generated, whose rows are the vectors $\Vec{x}_{*j}$. As $\Vec{y}_*$ (a column vector of length $m$) is correlated to $\Vec{y}$, they are drawn from the same multivariate Gaussian: 
\begin{equation*}
    \begin{bmatrix}
        \Vec{y}\\
        \Vec{y_*}
    \end{bmatrix} 
    \sim \mathcal{N}\left(\Vec{0},
    \begin{bmatrix}
        K & K_*\\
        K_*^T & K_{**}
    \end{bmatrix}
    \right),
\end{equation*}
where $K_*=\kappa(X,X_*) \text{ and } K_{**}=\kappa(X_*,X_*)$. Essentially, the known datapoints can be thought of as one sample drawn from this multivariate distribution. 

By using the conditional of a multivariate Gaussian, a prediction for $\Vec{y}_*$ \cite[Theorem 4.3.1]{murphy} can be obtained: 
\begin{equation*}
    P(\Vec{y}_*\mid X,X_*,\Vec{y})\sim \mathcal{N}(\Vec{\mu}_*,\Sigma_*),
\end{equation*}
where
\begin{align*}
    \Vec{\mu}_*&=K_*^T K^{-1}\Vec{y},\\
    \Sigma_*&=K_{**}-K_*^TK^{-1}K_*.
\end{align*}

Thus, the GP now has a prediction for both the mean and covariance matrix and thus the standard deviation, of $\Vec{y}_*$.

\section{Specifics of this Model}\label{sec:model_specifics}
The section highlights features specific to the Gaussian Process model demonstrated in this paper. All of the code is available on GitHub at the public repository~\cite{gp_git}. This model has been developed to work with pseudoscalar meson photoproduction data taken by the CLAS Collaboration based at Jefferson Lab, Virginia, USA, for the purposes of studying baryon spectroscopy, but can in principle be used on any dataset. The observables are defined in terms of energy and scattering angle.

\subsection{Choice of Kernel}
In the model demonstrated here, the radial basis function (RBF) \cite{sklearn} is used as the kernel function: 
 \begin{equation*}
    \kappa(\Vec{a},\Vec{b},\Vec{l})=\exp \left( \sum_{k=0}^{p-1} \frac{-\lVert (a_k-b_k)\rVert ^2}{2l_k^2}\right),
\end{equation*}
where $\Vec{l}$ is a hyperparameter called the length scale. In the case of the RBF kernel, this is related to the smoothness of the posterior distribution.

\subsection{Determining the Hyperparameters using Bayesian Inference}
Determining the correct values of hyperparameters is crucial to obtaining an acceptable fit. The standard approach is to use the log marginal likelihood as defined by Rasmussen and Williams in \cite{rasmussen_gp}:

\begin{equation}
\log P(\Vec{y} \mid X, \Vec{l}) = \frac{1}{2} \mathbf{y}^T K^{-1} \mathbf{y} 
+ \frac{1}{2} \log |K| + \frac{n}{2} \log 2\pi. \label{eq:log_marg}
\end{equation}

However, this is not well-defined for sparse datasets with a low number of datapoints, often leading to unrealistically small length scales which produce unphysical results. A different methodology must be employed to produce a reasonable fit when dealing with sparse datasets, as is often the case in hadron spectroscopy.

\begin{figure}[H]
    \centering
    \includegraphics[width=1.0\columnwidth]{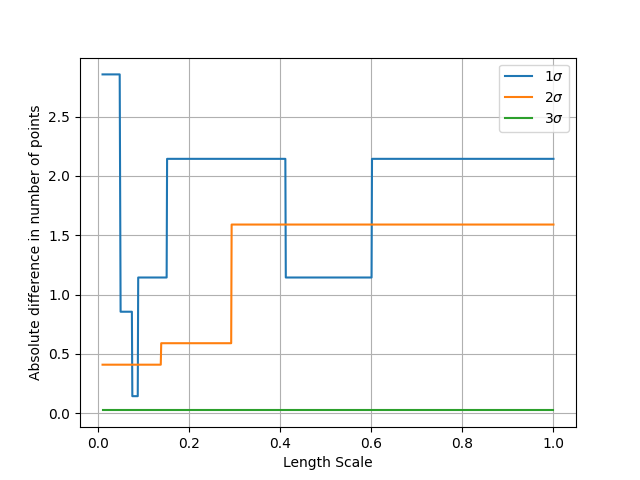} % Adjust size as needed
    \caption{Different $\sigma$ curves showing the difference between the expected number of datapoints and the measured number of datapoints for different length scales.}
    \label{fig:diff_sigma}
\end{figure}

For the purposes of a simpler explanation, assume a 1D dataset. From standard statistics, it is known that 68.3\% of the datapoints should be within one standard deviation of the fitted mean. The GP can return values and uncertainties using different values for the length scale hyperparameter, $l$; as $l$ is varied, the number of datapoints within one GP standard deviation of the GP mean also varies. The difference between the expected number of datapoints within 1 standard deviation (68.3\% of the total) and the measured number can be computed. Additionally, the difference between the expected number of datapoints with 2 and 3 standard deviations (95.5\% and 99.7\% respectively) compared to the measured number can also be computed. These differences are shown in Figure \ref{fig:diff_sigma}.
and are of the form: 
\begin{equation*}
    f(\vec{l}) = n |M(c\sigma,\vec{l}) - P(c\sigma)|
\end{equation*}
where 
\begin{itemize}
    \item $M(c\sigma, \vec{l})$ is the measured percentage of datapoints falling within $c\sigma$ of the GP mean for a given length scale $\vec{l}$,
    \item $P(c\sigma)$ is the corresponding predicated percentage.
    \item $n$ is the total number of datapoints.
\end{itemize}

The step-like appearance of the $1\sigma$ curve is a consequence of the limited number of datapoints; as the dataset size increases, the curve becomes smoother and more continuous. Furthermore, as the $3\sigma$ curve is flat, it shows that all of the length scales give GP fits with good coverage at high $\sigma$. 

Consequently, the length scale which has a minimum in the $1\sigma$ is also one which has the ``minimum'' in the 2 and $3\sigma$ curves. Therefore one can say this is the length scale which best matches the corresponding GP fit to the expectations from standard statistics. By taking the sum of these curves, this minimum can be found. However the choice of $c=1,2,3$, while intuitive, is somewhat arbitrary; any value from 0 to $3\sigma$ should also be considered. 

To avoid bias from selecting specific confidence intervals, the method is extended by marginalising over a continuous range of $c$ values from 0 to 3. As stated previously, confidence intervals up to $3\sigma$ cover approximately 99.7\% of data, thus offering a comprehensive evaluation range. By computing the loss over 1000 evenly spaced values of $c$ in the interval $(0,3]$, the summation approximates an integral over the full range of interest:

\begin{equation}
    f(\vec{l}) = \int_0^3  |M(c\sigma,\vec{l}) - P(c\sigma)|\, dc \approx \sum_{c>0}^{3} |M(c\sigma,\vec{l}) - P(c\sigma)|,\label{eq:loss_func}
\end{equation}

\begin{figure}[H]
    \centering
    \includegraphics[width=1.0\columnwidth]{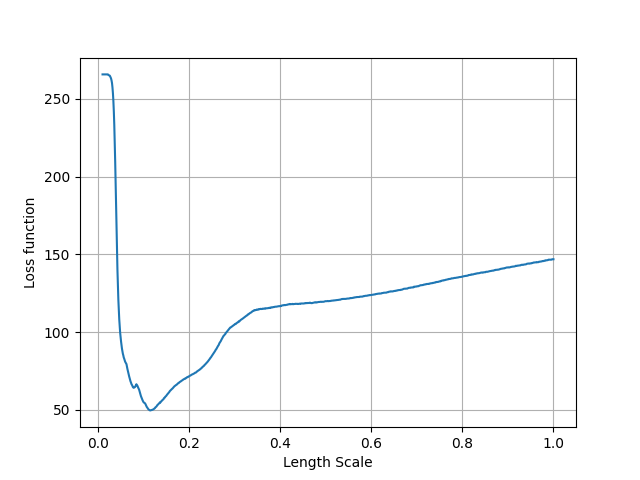} % Adjust size as needed
    \caption{An example loss function distribution.}
    \label{fig:loss_func}
\end{figure}

This yields a length scale loss function that captures the total statistical deviation between the GP model and expected coverage across all standard deviation ranges — offering a more holistic and statistically meaningful basis for hyperparameter selection in sparse data regimes. An example distribution of the loss function from some pseudodata (see Section \ref{sec:testing}) is shown in Figure \ref{fig:loss_func}.

For observables dependent on one kinematic variable, it is sufficient to use a standard minimiser to find the optimal length scale. However, increasing the kinematic dimensions, to even 2D, produces a multi-modal surface for the loss function; to solve this, a Markov Chain Monte Carlo (MCMC) can be used. 

In the MCMC procedure outlined in this paper, the chain is run until two different convergence criteria are met:
\begin{itemize}
    \item Gelman-Rubin Statistic \cite{gelman}, $\hat{R}<1.18$, (see Appendix \ref{sec:r_hat}).
    \item Integrated autocorrelation time stability \cite{emcee_paper}, $\tau_{stability}<0.15$, (see Appendix \ref{sec:tau_stability}).
\end{itemize}
Rather than running the MCMC for a pre-determined number of steps, convergence criteria can be checked regularly and will end the sampling when these criteria are met. Note that these criteria are looser than most general cases (usually $\hat{R}<1.1, \tau_{stability}<0.1$ are required). This is justified as the loss surface is often multi-modal, and the modes can have a similar value of $f(\vec{l})$, (equation \ref{eq:loss_func}). Stricter convergence criteria tend to return similar results but with significantly increased computational time. An example 2D corner plot produced by running the MCMC on some pseudodata (explained in section \ref{sec:testing}) with the above convergence criteria is shown in Figure \ref{fig:corner_plot} below.

\begin{figure}[H]
    \centering
    \includegraphics[width=1.0\columnwidth]{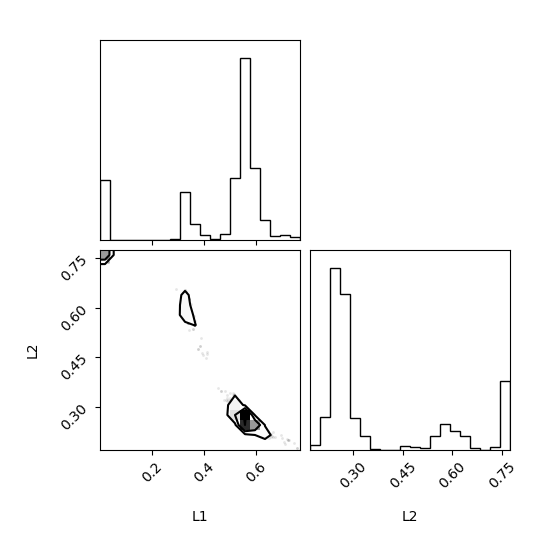} % Adjust size as needed
    \caption{An example 2D corner plot showing the multimodal nature of the loss function distribution.}
    \label{fig:corner_plot}
\end{figure}

To find the optimal length scale from this, the MCMC samples can be run through a K-means clustering algorithm, using silhouette scoring (first proposed by Rousseeuw in \cite{silhouette_scoring}), to determine the appropriate number of clusters. As K-means clustering only works for $K\ge2$, it must first be determined if there is only a single cluster, which can be achieved using \texttt{scipy.signal.find\_peaks} \cite{scipy}. If this finds one peak, the optimal length scale has been found and the algorithm ends. If more than one peak is found, silhouette scoring is used to determine the optimal $K$ by testing a range of possible $K$ values. 

Given $K$ clusters, the algorithm begins by selecting $K$ random MCMC samples to be the initial cluster means. Each MCMC sample is then assigned to its nearest cluster, which minimises the Euclidean distance. The means of the clusters are updated by taking the mean of all datapoints assigned to each cluster. Cluster assignment and cluster mean updating are repeated until convergence or the maximum number of iterations is reached; for \texttt{sklearn.clusters.KMeans} \cite{sklearn} the algorithm converges when the means of the clusters change by less than $10^{-4}$, or after 300 iterations by default. 

For these $K$ clusters, the silhouette score can be calculated. For each datapoint $x$, compute the mean distance between it and every other datapoint in its cluster, denoted as $a(x)$. Similarly, compute the mean distance between $x$ and every datapoint in the next nearest cluster, denoted as $b(x)$. From this the silhouette score for $x$ can then be defined as:
\begin{equation}
    S(x) = \frac{b(x)-a(x)}{\max(a(x),b(x))}\label{eq:silhouette_score}
\end{equation}
and the mean of these silhouette scores for all $x$ determines the overall silhouette score for $K$ clusters. 

This process is repeated for multiple $K$ values until the one that maximises the silhouette score is found. The means of the clusters, using the optimal $K$, are each used as initial guesses in a standard minimiser to determine the optimal length scale.

\subsection{Defining Points to Predict}\label{sec:convex_hull}
The GP presented here is intended to provide a data-driven interpolation of values and uncertainties for kinematic values that are within the domain defined by the measured datapoints. Testing has shown that the GP performs poorly when extrapolating beyond the range of the measured data; therefore, it is critical to ensure that any inputs to the GP are within a valid interpolation domain.

For 1D, this domain is simply defined as the interval bounded by the highest and lowest values of kinematic variable. In higher dimensions, the interpolation domain is defined by the convex hull enclosing the measured data points. The convex hull is the smallest convex set in Euclidean space that contains all the measured datapoints. A set \textit{S} is convex if, for any two points $a,b\in S$, the entire straight line segment connecting them is also in \textit{S} \cite{convex_hull}. The convex hull is used to define $X_*$, the matrix of input vectors for which the corresponding scalar values are interpolated. 

However, it is often the case that known datapoints of hadron spectra will be bins whose outer edges should be taken into account when defining the convex hull ($X_*$) of the known datapoints (\textit{X}). In the cases where the binning information is provided, this can easily be incorporated into the procedure. Assuming that the binning information is not available, the following procedure is used: 
\begin{itemize}
    \item Angle: The angular bins are only relevant in the context of creating the convex hull at the lowest (leftmost) and highest (rightmost) angular points. The widths of both the leftmost and rightmost bins are defined as half the distance to the nearest degenerate datapoint. Assuming \textit{m} measured degenerate datapoints, the angle bin width $\Delta(\cos\theta)$ is expressed mathematically as: 
\begin{equation*}
    \Delta(\cos\theta)= 
\begin{cases}
    \frac{|a_2-a_1|}{2},& \text{if } i=1.\\
    \frac{|a_m-a_{m-1}|}{2},& \text{if } i=m.\\
\end{cases}
\end{equation*}
    
    \item Energy: If the energy value is the highest or lowest energy measured, the bin width is half the distance to the adjacent measured energy value. If not, then the bin width is half the mean distance to the next highest and next lowest energy values. Assuming \textit{n} measured energy values, the energy bin width $\Delta E$ is expressed mathematically as: 
\begin{equation*}
        \Delta E= 
\begin{cases}
    \frac{|E_2-E_1|}{2},& \text{if } i=1.\\
    \frac{|E_n-E_{n-1}|}{2},& \text{if } i=n.\\
    \frac{1}{2}\left( \frac{|E_i-E_{i-1}|}{2}+\frac{|E_{i+1}-E_i|}{2}\right),              & \text{if } 2\leq i\leq n-1.
\end{cases}
\end{equation*}
\end{itemize}

In the scenario that there is no bin information available and the above conditions are not met (for example only one energy value is measured, each energy value only has one datapoint), the bin edges will be decided on a case-by-case basis. 

The above procedure has been designed with 2D pseudoscalar meson photoproduction in mind. For other data (different dimensions or kinematic dimensions) a similar procedure should be developed. 

\section{Testing Pseudodata}\label{sec:testing}
The pseudodata tested here is based on polarisation observables of the $\gamma + p \rightarrow K^0 + \Sigma^+$ reaction \cite{l_clark}, whose angular-dependent parts are sums of Legendre polynomials. It was assumed for testing purposes that the energy-dependent part consists of Gaussians, which are dependent on the order of Legendre polynomial. As such the 2D pseudodata surfaces are of the form: 
\begin{equation}\label{eq:func_form}
    f(w,\cos\theta)=\sum_{l=0}^{n}c_l g_l(w) P_l(\cos\theta),
\end{equation}
\begin{align*}
    \text{where } &c_l \in [-1,1] \text{ is some random weight}\\
    &g_l(w)\sim \mathcal{N}(\mu_l,\sigma_l^2) \text{ are Gaussians}\\
    &P_l(\cos\theta) \text{ is the }l\text{-th order Legendre polynomial}
\end{align*}

To evaluate the GP model's performance, 100 pseudodata surfaces are generated with $n=4$ (a typical expectation for baryon resonance analysis). From each surface, $p$ random points are selected to be the GP input. To best emulate real-world data, these $p$ points are chosen such that they span four discrete energy values, with each level containing between 4 and 10 datapoints. Noise and corresponding error bars are added to these points to reflect measurement uncertainty. Once the GP has been run for pseudodata, a set of $p$ random points is drawn from the GP fit to ensure similar coverage of the kinematic space. Both the original pseudodata and the sampled GP datapoints are then analysed to verify that the generated pseudodata and the result of the GP fitting procedures represent the same underlying information.

\subsection{Statistical Consistency}
We first verify that both the generated pseudodata and sampled GP datapoints exhibit the expected scatter about the underlying functions (equation \ref{eq:func_form}). The pull of the datapoints can be found using: 
\begin{equation*}\label{eq:pull}
    \text{pull}=\frac{z_{func}-z_{eval}}{\delta z_{eval}},
\end{equation*}
and rearranged to give:
\begin{equation*}
    |\text{pull}|\leq 1 \implies z_{func} \in [z_{eval}-\delta z_{eval},z_{eval}+\delta z_{eval}]
\end{equation*}
where $z_{eval}$ is either the pseudodata mean or the sampled GP mean, depending on what is being tested. Similarly $\delta z_{eval}$ is either the pseudodata uncertainty or the sampled GP uncertainty. In either case, $z_{func}$ is the true function. By setting $\delta z_{eval}$ to different multiples of the standard deviation (e.g. $0.67\sigma, 1\sigma,$ etc.) and calculating the resulting pull, the number of points in different multiples of the standard deviation can be calculated. This is repeated for each generated surface and the resulting mean number of points in each $\sigma$ band calculated, shown in Table \ref{tbl:sigma}; from this, it can be seen that both the pseudodata and sampled GP datapoints show pull distributions consistent with statistical expectations.

\begin{table}[H]
\begin{center}
\begin{tabularx}{1.0\columnwidth} { 
  | >{\arraybackslash}X 
  | >{\arraybackslash}X 
  | >{\arraybackslash}X 
  | >{\arraybackslash}X | }
\hline
$\sigma$ Band & Expected Percentage of Points (\%) & Measured Percentage of Known Points (\%) & Measured Percentage of GP Points (\%)\\
\hline
$0.67\sigma$& 50.0 & 51.2 & 59.0 \\
\hline
$1\sigma$& 68.3 & 68.1 & 73.1 \\ 
\hline
$1.96\sigma$& 95.0 & 94.8 & 94.4 \\ 
\hline
\end{tabularx}
\end{center}
\caption{Different $\sigma$ bands and the measured percentage of points contained in each}
\label{tbl:sigma}
\end{table}

\subsection{Fitting the Functional Form}\label{sec:fitting_func}
\begin{figure}[H]
    \centering
    \begin{subfigure}{1.0\columnwidth}
        \includegraphics[width=\columnwidth]{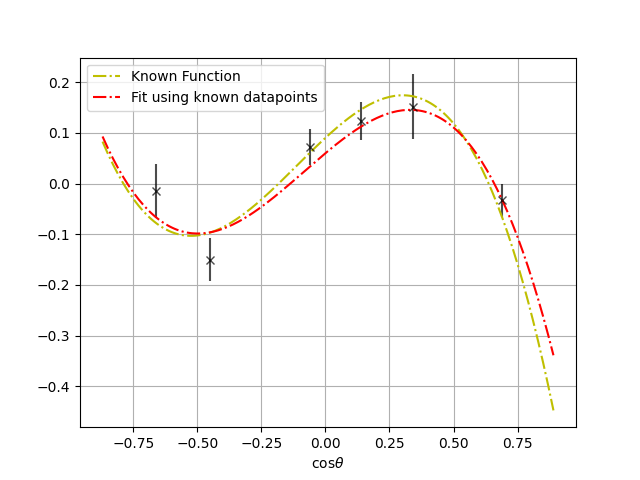}
        \caption{Fit to functional form using pseudodata}\label{fig:known_fit}
    \end{subfigure}

    \begin{subfigure}{1.0\columnwidth}
        \includegraphics[width=\columnwidth]{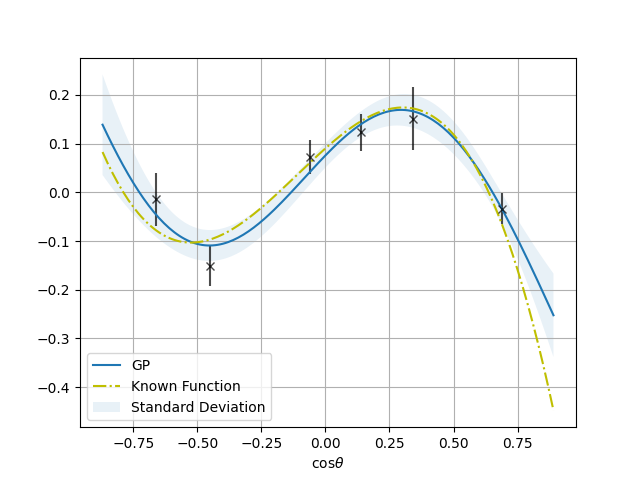}
        \caption{GP fit from pseudodata}\label{fig:GP_fit}
    \end{subfigure}

    \begin{subfigure}{1.0\columnwidth}
        \includegraphics[width=\columnwidth]{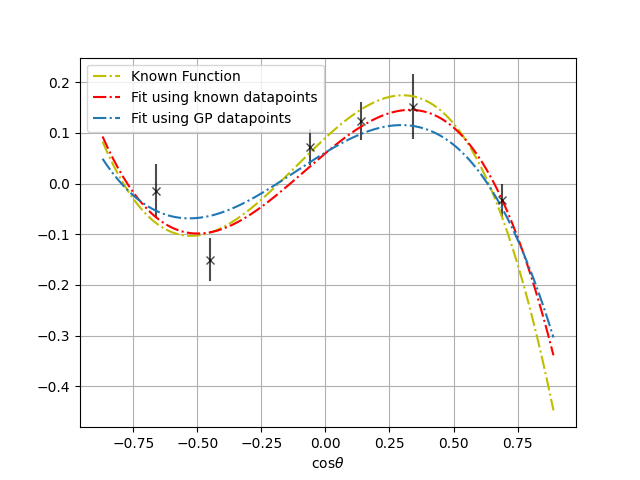}
        \caption{Fits to functional form using pseudodata and sampled GP datapoints}\label{fig:GP_coeff_fit}
    \end{subfigure}
    \caption{1D projections of fits from a generated 2D surface}
\end{figure}

By using the \texttt{Minuit} software package \cite{iminuit}, the functional form of the 2D surface, shown in Equation \ref{eq:func_form}, can be fitted to the known datapoints, using a least squares method, which is shown in Figure \ref{fig:known_fit}. A Gaussian Process fit is then performed on the same datapoints (Figure \ref{fig:GP_fit}) and the GP datapoints are used to fit the functional form of the 2D surface. The resulting fit, highlighted in Figure \ref{fig:GP_coeff_fit}, shows that fitting the functional form using the ``known'' datapoints or the GP datapoints produces similar results.

This can be further verified by finding the pull distribution of each of the surface coefficients which should be Gaussians centred at 0 with variance 1. The results are shown in Table \ref{tbl:coeff_pulls}; from this, it can be seen that both the pseudodata and sampled GP datapoints show coefficient pull distributions consistent with statistical expectations.

\begin{table}[H]
\begin{center}
\begin{tabularx}{1.0\columnwidth}{|>{\centering\arraybackslash}X|>{\raggedleft\arraybackslash}X|>{\raggedleft\arraybackslash}X|>{\raggedleft\arraybackslash}X|>{\raggedleft\arraybackslash}X|}
\hline&\multicolumn{2}{|c|}{Pseudodata} & \multicolumn{2}{|c|}{GP Datapoints}\\
\hline
Coefficient&Mean&Variance&Mean&Variance\\
\hline
$c_0$&0.04&0.88&0.08&0.92\\
\hline
$\mu_0$&0.07&0.81&0.14&0.51\\
\hline
$\sigma_0^2$&-0.12&0.74&-0.11&0.91\\
\hline
$c_1$&0.10&0.97&0.12&0.91\\
\hline
$\mu_1$&-0.12&0.55&-0.12&0.58\\
\hline
$\sigma_1^2$&-0.30&1.00&-0.13&0.79\\
\hline
$c_2$&0.11&0.99&-0.03&0.56\\
\hline
$\mu_2$&0.07&0.41&0.04&0.50\\
\hline
$\sigma_2^2$&-0.02&0.80&0.02&0.65\\
\hline
$c_3$&0.11&0.95&0.05&0.83\\
\hline
$\mu_3$&0.15&0.56&-0.13&0.58\\
\hline
$\sigma_3^2$&-0.25&0.87&-0.26&0.68\\
\hline
\end{tabularx}
\end{center}
\caption{Means and Variances of Pulls of fitted coefficients}
\label{tbl:coeff_pulls}
\end{table}

\section{Performing a GP fit on real data}\label{sec:GP_real_data}
Having shown that the GP methodology works for 2D polarisation observable pseudodata, by demonstrating that the two tests outlined in Section \ref{sec:testing} have been satisfied, the method can now be applied to measured data. As an example, we examine the photon beam asymmetry $\Sigma$ from the $\gamma + p \rightarrow K_0 + \Sigma^+$ reaction, as measured by the CLAS collaboration \cite{l_clark}.  The results are shown in Figures \ref{fig:sigma_2d_all} and \ref{fig:sigma_1d_proj} below. 

\begin{figure}[H]
    \centering
        \includegraphics[width=\linewidth]{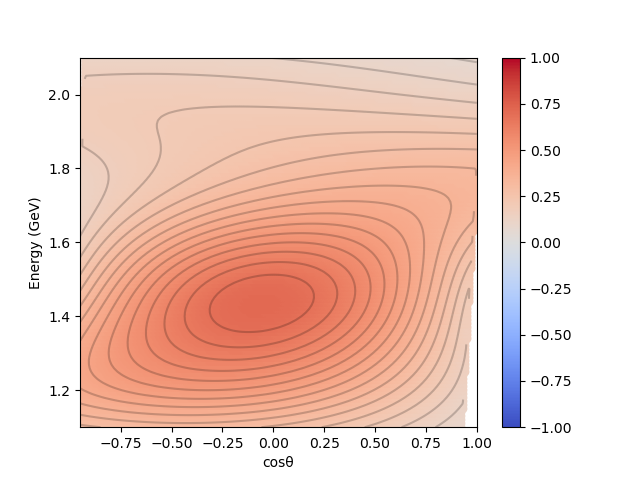}
        \label{fig:sigma_2d_surface}
    \caption{2D Contour plot of GP fit of $\Sigma$}\label{fig:sigma_2d_all}
\end{figure}

\begin{figure}[H]
    \centering
    \begin{subfigure}{1.0\columnwidth}
        \includegraphics[width=\linewidth]{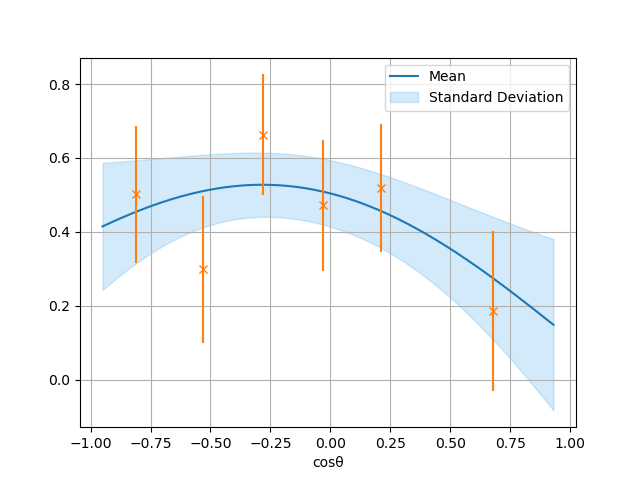}
        \subcaption{1D projection of GP fit at 1.23GeV.}
        \label{fig:sigma_E0}
    \end{subfigure}
%\end{figure}
%\begin{figure}[H]
    \centering
    %\ContinuedFloat
    \begin{subfigure}{1.0\columnwidth}
        \includegraphics[width=\linewidth]{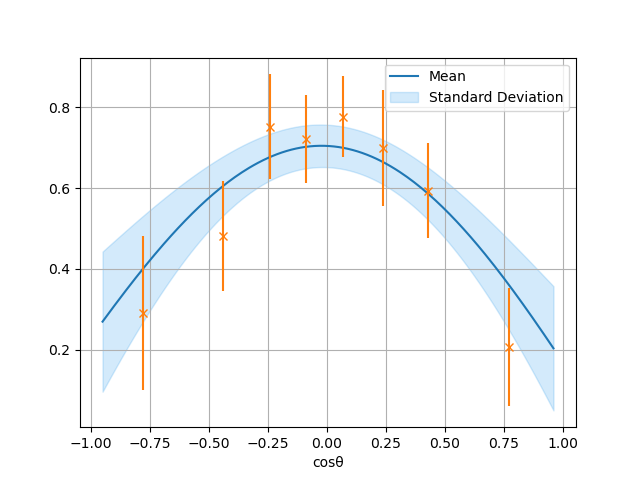}
        \subcaption{1D projection of GP fit at 1.48GeV.}
        \label{fig:sigma_E1}
    \end{subfigure}
% \end{figure}
% \begin{figure}[H]
    \centering
    % \ContinuedFloat
    \begin{subfigure}{1.0\columnwidth}
        \includegraphics[width=\linewidth]{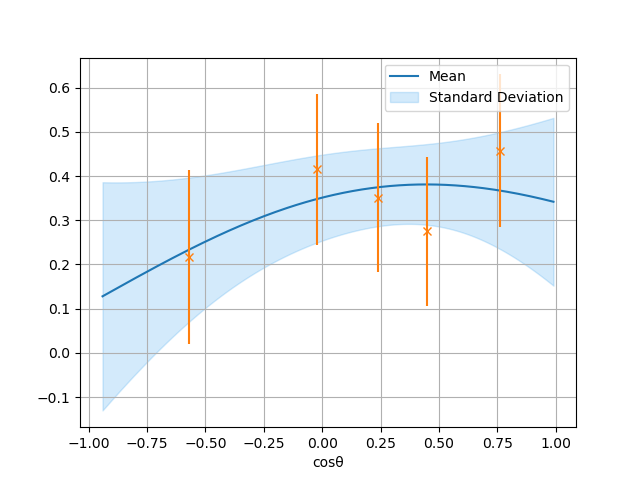}
        \subcaption{1D projection of GP fit at 1.73GeV.}
        \label{fig:sigma_E2}
    \end{subfigure}
 \end{figure}
 \begin{figure}[H]
    \centering
     \ContinuedFloat
    \begin{subfigure}{1.0\columnwidth}
        \includegraphics[width=\linewidth]{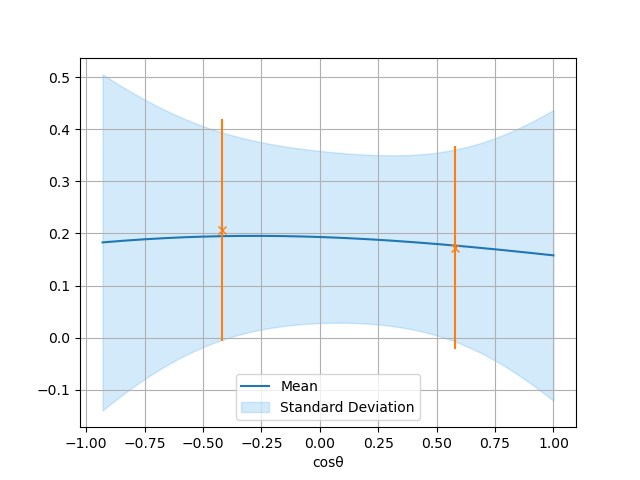}
        \subcaption{1D projection of GP fit at 1.98GeV.}
        \label{fig:sigma_E3}
    \end{subfigure}
    \caption{1D projections of GP fit of $\Sigma$}\label{fig:sigma_1d_proj}
\end{figure}

The usefulness of this method is that all measured observables required for a fit by a theoretical model can be treated in the same way; the GP method can return a value and uncertainty for an observable at any point as a function of kinematic variables (provided that it is an interpolation within the kinematic range of the measured data, see Section \ref{sec:convex_hull}).

This could be particularly useful in fits of coupled-channels models to data, which necessarily require measured data from several experimental sources. Data to be used in fits can be chosen with the same density of points in kinematic space, thereby removing the need for any arbitrary relative weighting of different observables that is sometimes required when different measured data sets consist of very different numbers of points.

\section{Data consistency}\label{sec:data_consistency}
\begin{figure}[H]
    \centering
    \includegraphics[width=1.0\columnwidth]{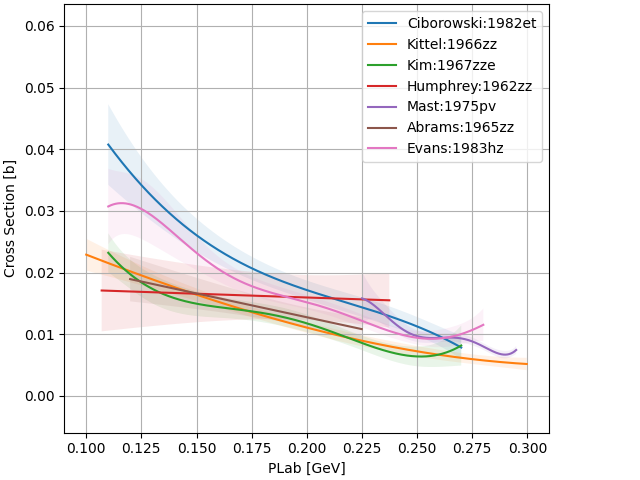} 
    \caption{Individual GP fits on experiments for $K^{-}p\rightarrow\bar{K}^{0}n$.}
    \label{fig:all_exps}
\end{figure}

Using this method, potential inconsistencies between different experiments that measure the same physics quantity can also be identified. As another example from hadron resonance physics, we examine the world data for $K^{-}p\rightarrow\bar{K}^{0}n$, 
\cite{Ciborowski,Kittel,Kim,Humphrey,Mast,Abrams,Evans}. A GP fit was run on each of the seven individual experiment. The graph in Figure \ref{fig:all_exps} shows the results of the seven GP fits (mean value and one standard deviation uncertainty band) plotted in different colours. It is clear from this that there is significant variation in the estimated cross section values, particularly at lower energy. 

This is a case where the standard procedure of combining estimates and uncertainties to give a single estimate as a function of energy might lead to misleading results. We can use the results of the GP fits to construct a combined probability ``surface'', which in this case is clearly not going to be a simple Gaussian form in the vertical direction.

\begin{figure}[H]
    \centering
    \includegraphics[width=1.0\columnwidth]{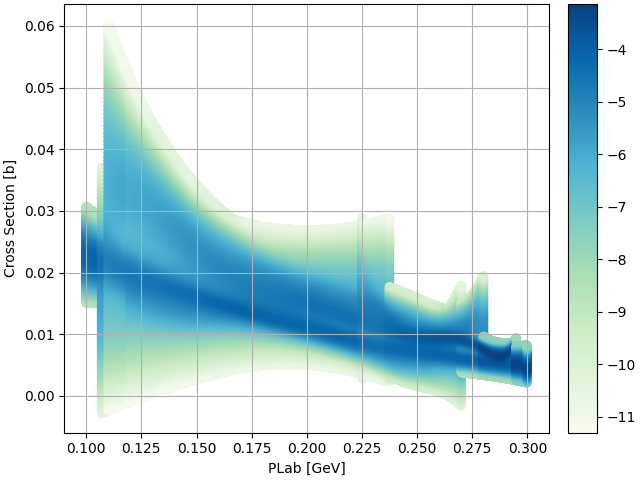} 
    \caption{Log probability surface for $K^{-}p\rightarrow\bar{K}^{0}n$.}
    \label{fig:log_prob_surface}
\end{figure}

For a given energy value, assume that each experiment with a corresponding GP measurement forms a normalised Gaussian and then take the sum of these Gaussians; this ensures that GP fits with a large uncertainty will have a small amplitude and thus will contribute less to the overall sum. Normalising this combined curve gives the probability density for this energy. To determine the probability of a specific cross section value, we need to multiply the probability density by some small $\delta y$. By repeating this for every energy value with a GP fit, the (log) probability surface for this channel can be found, as shown in Figure \ref{fig:log_prob_surface}.

It is clear from the plot that there is significant non-trivial structure in the likelihood surface, with a clear double ridged structure at lower energies. A theoretical fit that minimises least squares will attempt to move to values that represent the mean values. This will likely fall between ridges of higher probability and so will not be the maximum likelihood, thus producing misleading inferences.

This method provides a systematic, uncertainty-aware approach of comparing experimental datasets of the same physics quantity. By constructing a (log) probability surface from individual GP fits, regions of high and low agreement between experiments can be identified, with areas of significant disagreement indicating potential systematic discrepancies or unaccounted-for effects. This allows for a more rigorous assessment of dataset consistency ensuring agreement across multiple datasets, something essential for accurate theoretical modelling.

\section{Conclusion}\label{sec:conculsion}
The study of hadron resonances is fundamentally constrained by the accuracy, consistency and completeness of experimental datasets. In particular, coupled-channels analyses suffer from datasets with different coverages and binning, making direct comparison and theoretical fits challenging. Traditional methods of fitting to data often rely on arbitrary weighting schemes, which can introduce bias and remove any meaningful physical insights. 

This paper demonstrates that Gaussian Processes (GPs) provide a powerful, model-independent approach to overcoming these challenges. GPs enable the interpolation of sparse datasets with quantification of uncertainties. The use of Bayesian inference for hyperparameter optimisation ensures that the GP prediction adapts effectively to the underlying structure of the data. Verification using Legendre polynomials confirms this method's reliability, showing that GPs yield unbiased, accurate predictions. 

Unlike other machine learning methods that require arbitrary splitting of data into training and testing sets, GPs utilise the full dataset without the need for such partitioning. This ensures that all available experimental information contributes to the final prediction, reducing the risk of bias and improving robustness.

Beyond interpolation, GPs offer a systematic way to compare datasets from different experiments, helping to identify inconsistencies and improve the overall dataset. This capability is particularly valuable in coupled-channel analyses, where different quantities must be combined despite variations in experimental coverage and binning. Furthermore, the improved datasets produced by GPs provide theorists with a more reliable foundation for model validation and refinement. 

Overall, GPs present a significant opportunity for hadron spectroscopy by removing reliance on arbitrary weighting, improving dataset consistency and enhancing uncertainty quantification. The GP's ability to generate high-quality, data-driven interpolation makes them a valuable tool for experimental and theoretical studies, paving the way for more precise and reliable hadron resonance analyses.

\section*{Acknowledgments}
This work was supported by the United Kingdom’s Science and Technology Facilities Council (STFC) from Grant No. ST/Y000315/1.

\appendix

\section{Gelman-Rubin Statistic}\label{sec:r_hat}
The Gelman-Rubin Statistic, $\hat{R}$, checks that the samples are drawn from the same posterior distribution, by comparing the variance between different chains and the variance within chains. Using the emcee package \cite{emcee_paper} and assuming $m$ chains (walkers) and $n$ steps, the means of the chains $\bar{\theta_i}$ are found, and from this the means of means (or grand mean) $\bar{\theta}$. The mean variance within chains can be calculated using:
\begin{equation*}
    W = \frac{1}{m} \sum_{i=1}^{m} s_i^2,\label{eq:w_var}
\end{equation*}
where $s_i^2$ is the variance of the $i-th$ chain. Thus, $W$ represents the stability of each chain. 
The variance between chains can be calculated by:
\begin{equation*}
    B = \frac{n}{m-1} \sum_{i=1}^{m} (\bar{\theta}_i - \bar{\theta})^2.\label{eq:b_var}
\end{equation*}
As Gelman et al. \cite{gelman} say this $n$-scaling needs to be included; the true variance of the posterior is unknown so is estimated using the variance of the means of the chains. Since the variance of each chain mean is an estimate of $n$ samples, it underestimates the true variance by a factor$\frac{1}{n}$. Therefore we need to scale by $n$ to account for this. Thus $B$ represents the convergence of the chains. 

The variance within chains and the variance between chains are combined to give the pooled variance estimate:
\begin{equation*}
    \hat{V}=\frac{n-1}{n}W+\frac{1}{n}B. \label{eq:v_var}
\end{equation*}
This is done to get a truer representation of the posterior variance. From this, the potential scale reduction factor is defined as:
\begin{equation}
    \hat{R}=\sqrt{\frac{\hat{V}}{W}}. \label{eq:r_hat}
\end{equation}
Here when $\hat{R}\approx 1,\ \hat{V}\approx W$ so $B$ is small meaning the chains have mixed well and have converged to the same posterior distribution. If $\hat{R}>>1$ then $B$ is large so the chains have not converged.

\section{Integrated Autocorrelation Time Stability}\label{sec:tau_stability}
The Integrated Autocorrelation Time stability ($\tau_{stability}$) checks that the samples estimated are reliable, appropriately sized and have sufficiently explored the parameter space. Assuming $m$ walkers and $n$ steps, for a given chain, calculate the mean $\bar{\theta}$ and variance $s^2$ of the values $\theta$. From this, define the autocorrelation time as:
\begin{equation*}
    C(t)=\frac{1}{n-t}\sum_{s=1}^{n-t} (\theta_s-\bar{\theta})(\theta_{s+t}-\bar{\theta}),\label{eq:norm_autocorr}
\end{equation*}
which gives the normalised equation as:
\begin{equation*}
    \rho(t)=\frac{C(t)}{C(0)}.
\end{equation*}
Thus the integrated autocorrelation time for each chain (walker) can be calculated from:
\begin{equation*}
    \tau_{w}=\frac{1}{2}+\sum_{t=1}^{T_w}\rho_w(t).\label{eq:autocorr_cw}
\end{equation*}

Here $T_w$ is the point where $\rho_w(t)$ becomes negligible or is dominated by noise; in the methodology presented here, this is set to zero as the chain lengths are short. The integrated autocorrelation time can be found across all chains:
\begin{equation*}
    \tau=\frac{1}{m}\sum_{w=1}^{m}\tau_w.\label{eq:autocorr_c}
\end{equation*}
And thus the $\tau$ stability can be classified as:
\begin{equation}
    \tau_{stability}=\frac{|\tau_{old}-\tau_{new}|}{\tau_{new}}.\label{eq:tau_stable}
\end{equation}
Note that this assumes one kinematic dimension for which a length scale parameter is required. For higher dimensions, the MCMC calculates $\hat{R}\text{ and }\tau_{stability}$ for each dimension independently and only exits when each dimension satisfies the convergence criteria.

\bibliography{reference}% Produces the bibliography via BibTeX.
%\printbibliography
\end{document}